\documentclass[12pt]{iopart}
\usepackage{graphicx}

\def\bra#1{\mathinner{\langle{#1}|}} 
\def\ket#1{\mathinner{|{#1}\rangle}}

\begin{document}

\title[Local in time master equations]{Local in time master equations with memory effects: Applicability and interpretation}

\author{E.-M. Laine, K. Luoma and J. Piilo}

\address{Turku Centre for Quantum Physics, Department of Physics and Astronomy, University of Turku, FI-20014 Turun yliopisto, Finland}
\ead{emelai@utu.fi, ktluom@utu.fi}
\begin{abstract}
Non-Markovian local in time master equations give a relatively simple way to describe the dynamics of open quantum systems with memory effects. Despite their simple form, there are still many misunderstandings related to the physical applicability and interpretation of these equations. Here we clarify these issues both in the case of quantum and classical master equations. We further introduce the concept of a classical non-Markov chain signified through negative jump rates in the chain configuration.
\end{abstract}


\section{Introduction}\label{intro}
The interaction of a quantum system with its environment causes the dynamics to be non-unitary: the evolution of the state of the open quantum system can no longer be described by the Schr\"odinger equation but a master equation is needed in order to characterize dissipation and decoherence effects. Since merely all realistic quantum systems are open, i.e., they are influenced by their surroundings, the description of  such dynamics is crucial for our understanding of the nature. Therefore, the fundamental issues of open quantum systems as well as their mathematical description have gained a large amount of interest in the recent years \cite{Breuer1,Barreiro,measure1,measure2, Rivas1, Wolf, Rivas2, Shabani, appl1,appl2,Kossakowski2,Andersson1,nmj5}. 

The standard approach to the dynamics of open quantum systems, which employs the concept of a quantum Markov process, was developed already in the seventies \cite{Lindblad, Kossakowski1} and has been successfully applied in modeling many quantum systems \cite{mark1,mark2}. For a quantum Markov process the dynamics is given by a semigroup of completely positive dynamical maps and the corresponding master equation describing the dynamics is in the Lindblad form. The simple structure of the master equation allows a straightforward treatment of the dynamics of the open system whenever memory effects play no role in the dynamics.

However, for many processes occurring in the nature the approximations allowing the Markovian treatment are not valid. For example, strong system-environment coupling, structured or finite reservoirs and low temperatures can give rise to pronounced memory effects in the dynamics of the open system and these systems require a more sophisticated non-Markovian treatment. The increasingly important role of non-Markovian processes has initiated significant steps towards a general consistent theory of non-Markovian quantum dynamics  \cite{measure1,measure2, Rivas1, Wolf, Rivas2} as well as achievements in the experimental detection and control of memory effects \cite{exp1, exp2}.

The vast interest in memory effects in quantum processes has led into many mathematical formulations of the dynamics beyond the Markov theory as well as a large variety of tools for solving the dynamics. Probably the most common ones of these methods apply memory kernel master equations \cite{Breuer1, Shabani, Nakajima, Zwanzig} or local in time master equations \cite{Breuer1, local1}. The local in time master equations give a mathematically rather simple method for dealing with memory effects and they have been applied to describe a variety of non-Markovian processes in nature \cite{appl1,appl2,appl3,appl4,appl5}. Despite the simple form, there are still many misunderstandings related to the physical applicability and interpretations of these equations. Here we pinpoint and clarify some of these misbeliefs about the local in time description and give some illustrative examples to further explain these points.

The structure of the paper is as follows. In section \ref{ch2} we introduce the formulation of non-Markovian dynamics in terms of local-in-time master equations and specify the connection between the quantum and classical cases. We further connect the local in time equation to a quantum jump description \cite{Piilo1, Piilo2}, allowing an interpretation for negative decay rates in the equation. In section \ref{ch3} we discuss three different points regarding the local in time description, which are often misunderstood, and give some illustrative examples to rebut these misbeliefs. Section \ref{ch4} introduces a description of a classical non-Markov chain, where memory effects play a crucial role.

\section{Local in time master equations} \label{ch2}

For a Markovian quantum process the dynamics is given by a semigroup of completely positive dynamical maps and the corresponding master equation for the reduced density matrix $\rho(t)$ describing the dynamics is in the Lindblad form
\begin{equation}\label{eq1}
\frac{d}{dt}\rho(t)=-i\left[H,\rho(t)\right]+\sum_i \gamma_i \left(L_i\rho(t)L_i^\dagger-\frac{1}{2}\{L_i^\dagger L_i,\rho(t)\}\right),
\end{equation}
where $H$ is a Hermitian operator, $\gamma_i$ are the decay rates, which are positive constants, and $L_i$ are the Lindblad operators.

The standard approach to the non-Markovian dynamics of open systems uses the Nakajima-Zwanzig projection operator technique \cite{Nakajima, Zwanzig} from which one obtains the memory kernel master equation
\begin{equation}\label{eq2}
\frac{d}{dt}\rho(t)=\int_0^t \mathcal{K}(t-s)\rho(s)ds,
\end{equation}
in which quantum memory effects are taken into account through the introduction of the memory kernel $\mathcal{K}(t)$. Since the master equation is not local in time due to the integration over the density matrix $\rho$, it is usually taken as granted that a master equation of this form guarantees the occurrence of memory effects in the dynamics. However, the form of a master equation is not in general unique, i.e., the dynamics can be described equivalently by another type of master equation, which is local in time \cite{memker}. More precisely, it was shown in \cite{Kossakowski2, Andersson1}, that under fairly general assumptions, the master equation of the form of equation (\ref{eq2}) can always be cast into a local in time form
\begin{equation*}
\frac{d}{dt}\rho(t)=\mathcal{L}(t)\rho(t),
\end{equation*}
in which no memory kernel occurs. Further, any time-local master equation can be cast in a Lindblad-like form of equation (\ref{eq1}), but with time dependent Lindblad operators as well as time dependent decay rates $\gamma_i(t)$ which may also become negative for some intervals of time \cite{nmj5, Andersson2}. Thus, non-Markovian dynamics can be written in the local in time form
\begin{eqnarray}
\frac{d}{dt}\rho(t)&=&-i\left[H(t),\rho(t)\right] \label{eq3} \\
&&+\sum_i \gamma_i(t) \left(L_i(t)\rho(t)L_i^\dagger(t)-\frac{1}{2}\{L_i^\dagger(t) L_i(t),\rho(t)\}\right),\nonumber
\end{eqnarray}
with temporarily negative decay rates $\gamma_i(t)$. Master equations of this form have been frequently applied in the context of non-Markovian systems, but there are still some misunderstandings related to their applicability. Before going into details of these common misunderstandings in section \ref{ch3} let us first introduce some aspects related to these equations, which will allow us to explain our arguments in the following sections.

\subsection{Quantum and classical master equations} \label{ch21}
The master equation (\ref{eq3}) describes a general quantum process, which reduces to a standard quantum Markov process whenever the decay rates $\gamma_i(t)$ are positive constants and the Lindblad operators are independent of time. Also, as a special case, one can obtain the classical rate equation under certain assumptions as we will show here.

Let us first assume that the Hamiltonian of the system is time-independent and it has the spectral decomposition $H=\sum_k E_k \ket{\psi_k}\bra{\psi_k}$, where $E_k$ are the eigenenergies and $\ket{\psi_k}$ the eigenvectors of the Hamiltonian. Now assume, that the system is described by a classical probability distribution instead of the quantum mechanical density matrix, i.e., the density matrix can be written as a diagonal matrix in the system eigenbasis $\rho(t)=\sum_k p_k(t) \ket{\psi_k}\bra{\psi_k}$, where $p_k(t)$ gives the occupation probability of the eigenstate $\ket{\psi_k}$ at time $t$. Now, if we assume, that the Lindblad operators $L_i$ in equation (\ref{eq3}) describe transitions from one eigenstate to another $L_i=L_{kl}=\ket{\psi_k}\bra{\psi_l}$ such that $l\neq k$ we obtain
\begin{eqnarray}
\frac{d}{dt}\rho(t)&=&-i\left[H,\rho(t)\right]\nonumber\\
&&+\sum_{k}\sum_{l\neq k} \gamma_{kl}(t) \left(L_{kl}\rho(t)L_{kl}^\dagger-\frac{1}{2}\{L_{kl}^\dagger L_{kl},\rho(t)\}\right)\nonumber\\
&=&\sum_{k}\sum_{l\neq k} \gamma_{kl}(t)\left(p_l(t)\ket{\psi_k}\bra{\psi_k}-p_l(t)\ket{\psi_l}\bra{\psi_l}\right)\nonumber\\
&=&\sum_{k}\ket{\psi_k}\bra{\psi_k} \sum_{l\neq k} \left(\gamma_{kl}(t)p_l(t)-\gamma_{lk}(t)p_k(t)\right),\nonumber
\end{eqnarray}
from which one obtains the rate equation
\begin{equation}\label{eq4}
\dot{p}_k(t)=\sum_{l\neq k}\left(\gamma_{kl}(t)p_l(t)-\gamma_{lk}(t)p_k(t)\right),
\end{equation}
which can also be cast into a more compact form
\begin{equation}\label{eq4b}
 \dot{\vec{P}}(t)=\vec{P}(t)^T\mathbf{Q}(t),
\end{equation}
where $\vec{P}(t)$ is the column vector with elements $p_k(t)$ and $\mathbf{Q}(t)$ is a matrix with elements $q_{kl}(t)=\gamma_{kl}(t)$ for $k\neq l$ and $q_{ll}=-\sum_{k\neq l}\gamma_{lk}$.Whenever the decay rates are positive constants equations (\ref{eq4}) and (\ref{eq4b}) describe the well known continuous in time Markov chain. However, if the decay rates can be temporarily negative these equations no longer describe a Markov chain but rather a classical non-Markovian process. In the following chapters we will study further such classical non-Markovian processes as well as clarify the interpretation of negative rates in the classical realm. Before that, let us study more the interpretation of negative decay rates in the quantum case.

\subsection{Master equations and jump descriptions} \label{ch22}
The state of a quantum system is described by a density matrix, which can also be seen as an ensemble of state vectors each of them having a classical probability of appearance. This view has given the starting point for the development of Monte Carlo simulation methods for Markovian \cite{jump1, jump2, jump3,jump4} and non-Markovian \cite{nmj5,Piilo1, Piilo2, nmj1,nmj2, nmj3,nmj4,nmj6, Luoma} open quantum systems in which the time evolution of each state vector in the ensemble contains a stochastic element. One of these methods for Markovian dynamics is the Monte Carlo wave function (MCWF) method which exploits quantum jumps \cite{jump1}.

The MCWF quantum jump method is probably the most commonly used Monte Carlo method for treating Markovian open systems whose dynamics is governed by the master equation in the Lindblad form of equation (\ref{eq1}). To unravel the master equation (\ref{eq1}) one has to generate an ensemble of stochastic state vector realizations whose deterministic and continuous time evolution is interrupted by randomly occurring quantum jumps. The average over the ensemble of stochastic realizations gives the reduced system at any given moment of time. One can write the density matrix in terms of the ensemble of state vectors as
\begin{equation}\label{eq5}
\rho(t)=\sum_\alpha \frac{N_\alpha(t)}{N}\ket{\psi_\alpha(t)}\bra{\psi_\alpha(t)},
\end{equation}
where $N_\alpha(t)$ is the number of ensemble members in the state $\ket{\psi_\alpha(t)}$ at time $t$ and $N$ is the total ensemble size.

The method proceeds in discrete time steps $\delta t$ and a single time step that takes us from time $t$ to $t + \delta t$. During the time step, a given state vector $\ket{\psi_\alpha(t)}$ evolves either in a deterministic way or performs a randomly occurring quantum jump. The deterministic evolution is given by the non-Hermitian effective Hamiltonian
\begin{equation}\label{eq6}
H_{\textrm{eff}}=H-\frac{i}{2}\sum_i\gamma_iL_i^\dagger L_i,
\end{equation}
where the operators are the Hermitian operator $H$ and the Lindblad operators $L_i$ in Eq.~(\ref{eq1}). The deterministic time-evolution by the effective Hamiltonian (\ref{eq6}) gives for a single time step the state
\begin{equation}\label{eq7}
\ket{\phi_\alpha(t+\delta t)}=\left(1-i H_{\textrm{eff}} \delta t\right)\ket{\psi_\alpha(t)},
\end{equation}
after which the state is renormalized. If, on the other hand, a quantum jump to channel $i$ occurs, the state vector changes in a discontinuous way
\begin{equation}\label{eq8}
\ket{\psi_\alpha(t)}\rightarrow \ket{\psi_\alpha(t+\delta t)}=\frac{L_i\ket{\psi_\alpha(t)}}{||L_i\ket{\psi_\alpha(t)}||}.
\end{equation}
The probability $p_\alpha^i$ for a state vector $\ket{\psi_\alpha}$ to have a quantum jump to channel $i$ is given by
\begin{equation} \label{eq9}
p_\alpha^i(t)=\gamma_i \delta t \bra{\psi_\alpha(t)}L_i^\dagger L_i\ket{\psi_\alpha(t)}.
\end{equation}

Now, let us consider the non-Markovian case. For simplicity, we assume the Lindblad operators in the local in time master equation (\ref{eq3}) to be time-independent. The difference to equation (\ref{eq1}) is that the time dependent decay rates get temporarily negative values. One can directly see from equation (\ref{eq9}) that whenever the decay rate is negative one cannot apply the MCWF quantum jump method, since one would end up with negative jump probabilities. However it is possible to develop another jump description for the non-Markovian case, inherently different from the Markovian one \cite{Piilo1,Piilo2, Piilo3}.

The deterministic evolution is equivalent to the Markovian case, i.e. given by equations (\ref{eq6}) and (\ref{eq7}) \cite{Piilo2}. Also, whenever the decay rate in channel $i$ is positive the corresponding jump can be produced as the Markovian one. However, when the decay rate in channel $i$ turns negative, memory effects start to play a role and the jump method described in equations (\ref{eq8}) and (\ref{eq9}) can no longer be used. Indeed, for a negative channel $i$ the direction of the jump process gets reversed
\begin{equation*}
\ket{\psi_{\alpha'}(t+\delta t)}\leftarrow\ket{\psi_\alpha(t)}=\frac{L_i\ket{\psi_{\alpha'}(t)}}{||L_i\ket{\psi_{\alpha'}(t)}||}
\end{equation*}
and the jump operator for the negative channel $i$ takes the form
\begin{equation*}
D_{\alpha\rightarrow\alpha'}^i(t)=\ket{\psi_{\alpha'}(t)}\bra{\psi_{\alpha}(t)},
\end{equation*}
where the source state of the jump is $\ket{\psi_\alpha(t)}=L_i\ket{\psi_{\alpha'}(t)}/||L_i\ket{\psi_{\alpha'}(t)}||$. Thus, the source and the target states of the jump get swapped as the decay rate becomes negative. Further the probability for a reversed jump for a given state vector $\ket{\psi_\alpha}$ is given by
\begin{equation}\label{eq10}
P_{\alpha\rightarrow\alpha'}^i=\frac{N_{\alpha'}(t)}{N_\alpha(t)}|\gamma_i(t)|\delta t \bra{\psi_{\alpha'}(t)}L_i^\dagger L_i\ket{\psi_{\alpha'}(t)}.
\end{equation}
Note that if there are no ensemble members in the target state, $N_{\alpha'}=0$, the jump probability is equal to zero.

\section{Misbeliefs about local in time equations} \label{ch3}
The local in time equations of the form given in equation (\ref{eq3}) with temporarily negative decay rates give a natural extension of the Markovian equation (\ref{eq1}) to the non-Markovian regime. The negative decay rates arise naturally, when no Markov approximation is made in the microscopic derivation of the master equation. They have been widely applied to non-Markovian problems in physics, but there are still many misunderstandings related to the applicability and interpretation of these equations. In this section we pinpoint some of these misunderstandings and through some simple examples clarify these points.

\subsection{Negative decay rates and positivity of the density matrix} \label{ch31}
A naturally arising question is whether the dynamics remains physical when the Markovian master equation (\ref{eq1}) is extended to the non-Markovian regime where temporarily negative decay rates arise. The physicality of the process refers to the question of whether the master equation produces physical states, when the input state is physical. In the quantum realm this boils down to the question, whether the resulting dynamical map is completely positive for each time $t$. The complete positivity of the  dynamical map is a more strict requirement than the positivity of the output state, i.e., whenever the map is completely positive the output states must be positive as well.

It is well known, that the Markovian master equation (\ref{eq1}) produces completely positive dynamical maps for all times. However this requires the positivity of the decay rates. When the decay rates are temporarily negative there exists no general proof, that the resulting maps would be completely positive. However, if further assumptions on the properties of the decay rates are made, the requirement of complete positivity is met. In order to further illustrate this statement let us consider a simple example of a two-level system with a single channel.

The general master equation for a single channel process of a two-level system with the Lindblad operator $\sigma_-=\ket{g}\bra{e}$, where $\{ \ket{e},\ket{g}\}$ is a system basis, is
\begin{equation}\label{eq11}
\frac{d}{dt}\rho(t)= \gamma(t) \left(\sigma_-\rho(t)\sigma_+ -\frac{1}{2}\{\sigma_+ \sigma_-,\rho(t)\}\right),
\end{equation}
where $\gamma(t)$ is an arbitrary function with temporarily negative values. For simplicity we take the system Hamiltonian $H=0$. Equation (\ref{eq11}) can be easily solved to give the dynamical maps describing the process for time $t$
\begin{eqnarray}
\rho(t)&=&\kappa(t)\rho_{ee}(0)\ket{e}\bra{e}+(1-\kappa(t)\rho_{ee}(0))\ket{g}\bra{g}\nonumber\\
&&+\kappa(t)^{1/2}(\rho_{eg}(0)\ket{e}\bra{g}+\rho_{ge}(0)\ket{g}\bra{e}),\label{eq12}
\end{eqnarray}
where $\kappa(t)=\exp\left[-\int_0^t\gamma(s) ds\right]$.
One can proof that the dynamical maps for different times $t$ given by equation (\ref{eq12}) are completely positive if and only if $\int_0^t\gamma(s)ds\geq0$ for all times $t$. Thus in order for the master equation (\ref{eq11}) to describe a physical process, the decay rate $\gamma(t)$ needs not to be positive but only it's integral has to be. Therefore, in general, master equations with negative decay rates do produce physical processes, when certain additional constraints are met.

\subsection{Negative decay rates and jump probabilities} \label{ch32}
In the previous section we demonstrated, that negative decay rates in a time local master equation can produce completely positive dynamical maps, i.e., that they do describe physical processes. A natural question arising is, how to interpret the negative decay rates. For the Markovian case, with positive rates, the interpretation is straightforward when one follows the line of thought of the jump unraveling of section \ref{ch22}; the decay rate gives the rate at which jumps from the source state to the target state of the corresponding channel occur and thus the jump probabilities given in equation (\ref{eq9}) are directly proportional to decay rate.

A misunderstanding regarding the interpretation of the negative decay rates occurs, when one tries to use the Markovian logic in the non-Markovian case: one quite easily makes the false conclusion, that the negative decay rate only reverses the direction of the process between the state vectors. However, one can quite easily figure out, that this is not enough: in the non-Markovain jump unraveling of section \ref{ch22} the direction of the process is reversed, but also the jump probabilities are different: As we can see in equation (\ref{eq10}), the jump probability is not only proportional to the absolute value of the decay rate, but also to the ratio of the occupation numbers $N_{\alpha'}(t)/N_\alpha(t)$ given by the state of the system at the time of the jump. In particular, the jump probability is proportional to the occupation number of the target state, which carries information about the state of the system at previous times. Thus, whenever the decay rate is negative one can define an effective rate for the reversed jumps given by
\begin{equation}\label{eq13}
\tilde{\gamma}_{i}^{\alpha \rightarrow\alpha'}(t)=|\gamma_i(t)|\frac{N_{\alpha'}(t)}{N_\alpha(t)},
\end{equation}
which depends on the state the system at time $t$ and therefore also on the earlier states of the system. The simple reversal of the jump process, without changing the jump probabilities would naturally not produce a non-Markovian process. The effective decay rates in equation (\ref{eq13}) however encode the history of the process and thus allow memory effects in the dynamics. In the following section we will discuss memory effects described by local in time equations in detail, but first we will explore the concept of effective decay rates in the classical case.

Assume we have a classical stochastic process described by the rate equation (\ref{eq4}). When the rates $\gamma_{kl}(t)$ are positive, the terms with a positive sign represent probability gain to the state $k$ and the terms with a negative sign represent loss. Let us now consider a gain term $\gamma_{kl}(t) p_l(t)$ and assume $\gamma_{kl}(t)<0$. Now, the term can be rewritten
\begin{equation*}
\gamma_{kl}(t) p_l(t)=-|\gamma_{kl}(t)| \frac{p_l(t)}{p_k(t)}p_k(t)=-\tilde{\gamma}_{kl}(t)p_k(t),
\end{equation*}
and we see that the gain term with the negative rate $\gamma_{kl}(t)$ actually describes a loss term with an effective positive rate $\tilde{\gamma}_{kl}=|\gamma_{kl}(t)| \frac{p_l(t)}{p_k(t)}$. Thus, also in the classical case, the negative decay rate describes a reversed process with an effective decay rate, which depends also on the earlier states of the system.

\subsection{Local in time equations and memory effects} \label{ch33}
The fact that equation (\ref{eq3}) is local in time, i.e., there is no integration over the past states of the system in the master equation often creates a misunderstanding that the equation can not describe memory effects. However, the classical counterpart of the equation immediately shows that this is not the case whenever there are negative decay rates: The rate equation with positive rates is a straightforward consequence of the Markov property. Thus, whenever one cannot write a rate equation with positive rates for the process, it cannot fulfill the Markov condition. Therefore, the process must be non-Markovian. The classical case acts as a simple demonstration of the fact that local in time equations with negative decay rates indeed describe memory effects.

To further demonstrate the memory effects induced by negative decay rates, let us study a simple example of a two-state system and compare the probability flow in the non-Markovian and corresponding Markovian cases. We denote the probability to be on state $1(2)$ at time $t$ by $p_{1(2)}(t)$. Let us first study a Markovian model, where  the rate equations are 
\begin{equation*}
\left\{ \begin{array}{ll}
\dot{p}_1=\gamma_2(t)p_2(t)-\gamma_1(t)p_1(t)\\
\dot{p}_2=\gamma_1(t)p_1(t)-\gamma_2(t)p_2(t)
  \end{array}\right.
\end{equation*}
and the decay rates are
\begin{equation*}
  \gamma_1(t)=\left\{ \begin{array}{ll} 
   \Gamma & 0\leq t < s_1\\
  0 & s_1\leq t < s_2 \\
  \Gamma & s_2 \leq t < s_3 \\
   \vdots &  
  \end{array}\right.
  \quad 
  \gamma_2(t)=\left\{ \begin{array}{ll} 
  0 & 0\leq t < s_1 \\
  \Gamma & s_1\leq t < s_2 \\
  0 & s_2 \leq t < s_3 \\
   \vdots &  
  \end{array}\right..
\end{equation*}
This is a process, where the direction of the probability flow changes periodically due to two alternating channels with opposite direction, but there are no memory effects.

\begin{figure}
  \includegraphics[width=0.5\textwidth]{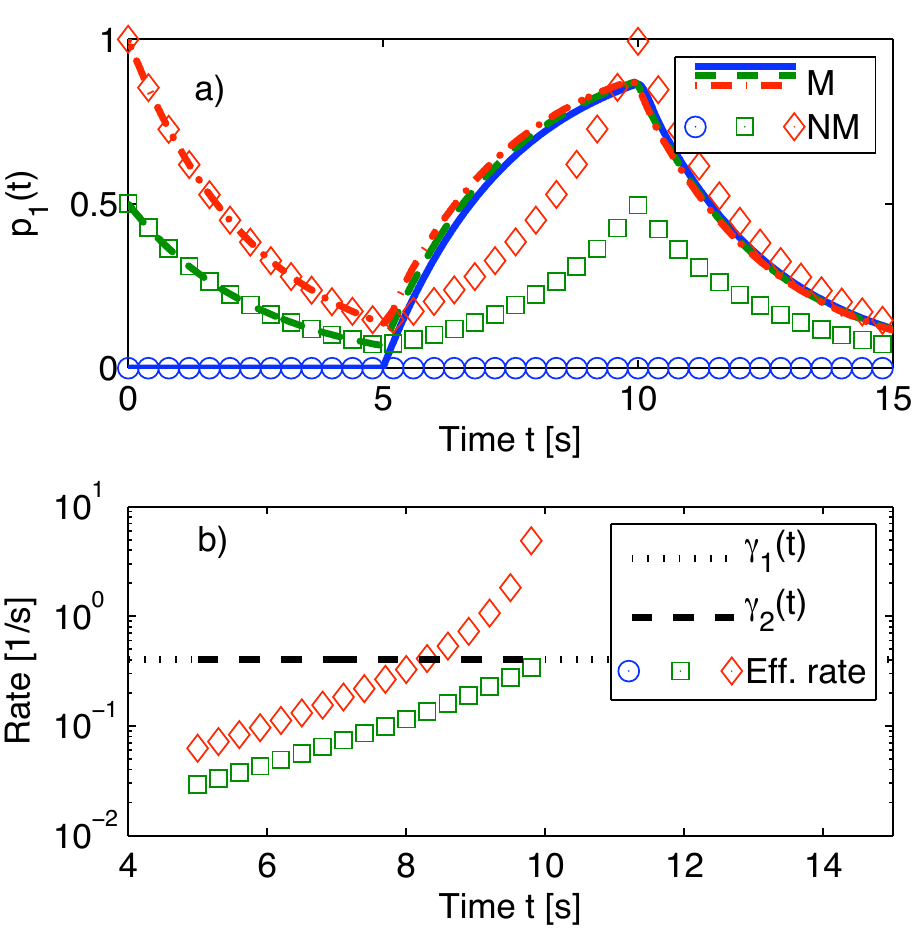}
  \caption{\label{fig:ex5} Markovian (M) and Non-Markovian (NM) 
    two state model for
    different initial states $p_1(0)={1,0.5,0}$ and
    for $\Gamma=0.4 \, 1/s$. The upper panel describes state of the 
    system as a function of time. Solid lines are for the Markovian
    and markers for the Non-Markovian model.
    In the lower panel we plot the time dependent rates for the Markovian model (dashed 
    and dotted lines), the positive rate for the Non-Markovian model (dotted line)
    and the effective rate for the Non-Markovian model (markers). Color and 
    marker coding matches to upper panel. The effective rate for $p_1(0)=0$ (circles) is zero for all times and therefore not displayed in the plot.}\label{fig1}
\end{figure}

The corresponding process with memory effects has only one decay rate $\gamma(t)$ for the transitions $1\rightarrow 2$ with a negative region for $s_1\leq t < s_2$, i.e.,  $\gamma(t)=\gamma_1(t)-\gamma_2(t)$. As discussed in the previous section, for the period of time with the negative decay rate we can rewrite the equation in terms of an effective decay rate, which in this case is $\tilde{\gamma}_2(t)=\gamma_2(t)\frac{p_1(t)}{p_2(t)}$ and the rate equation reads
\begin{equation*}
\dot{p}_1=-\gamma(t)p_1(t)=\tilde{\gamma}_2(t)p_2(t)-\gamma_1(t)p_1(t)
\end{equation*}
In figure \ref{fig1} we have plotted $p_1(t)$ for both the processes as well as the effective decay rate $\tilde{\gamma}_2(t)$ for $\Gamma=0.4 \, 1/s$, $s_1=5\, s$ and $s_2=10\, s$. We can see a substantial qualitative difference in the Markovian and non-Markovian cases. Naturally both the processes coincide for $t\leq 5\, s$. In this region the different initial values start to approach a stationary value independent of the initial state. For $t\geq 5\, s$, the negative decay rate occurs for the non-Markovian process and the two processes diverge from each other. For the Markovian process probability distributions corresponding to different initial values still approach each other, since the decay rate is independent of the initial state. For the non-Markovian process, the effective decay rate $\tilde{\gamma}_2(t)$ depends on the earlier states of the system and therefore the different probability distributions, corresponding to different initial states, diverge from each other when the decay rate is negative and signify the presence of memory effects. In the non-Markovian case the evolution for $t\geq5\,s$ is a reflection of the history of the state: the system starts approaching the initial state by reversing the process which occurred in the past. In contrast, for the corresponding Markovian process the system does not account for the past but continues to approach a common stationary value independent of the initial state. The most striking difference between the two processes can be seen, when the initial state is chosen to be $p_1(0)=0$. For this initial value, in the Markovian case $p_1(t)$ starts increasing. However, for the non-Markovian process $p_1(t)=0$ for all times $t$, since the effective decay rate is zero for all times; one cannot reverse jumps that never occurred. In the next section we further explore different characteristics of classical non-Markovian processes described by local in time equations.

\section{A classical non-Markov chain} \label{ch4}
\begin{figure}[!h]
\includegraphics[width=8cm]{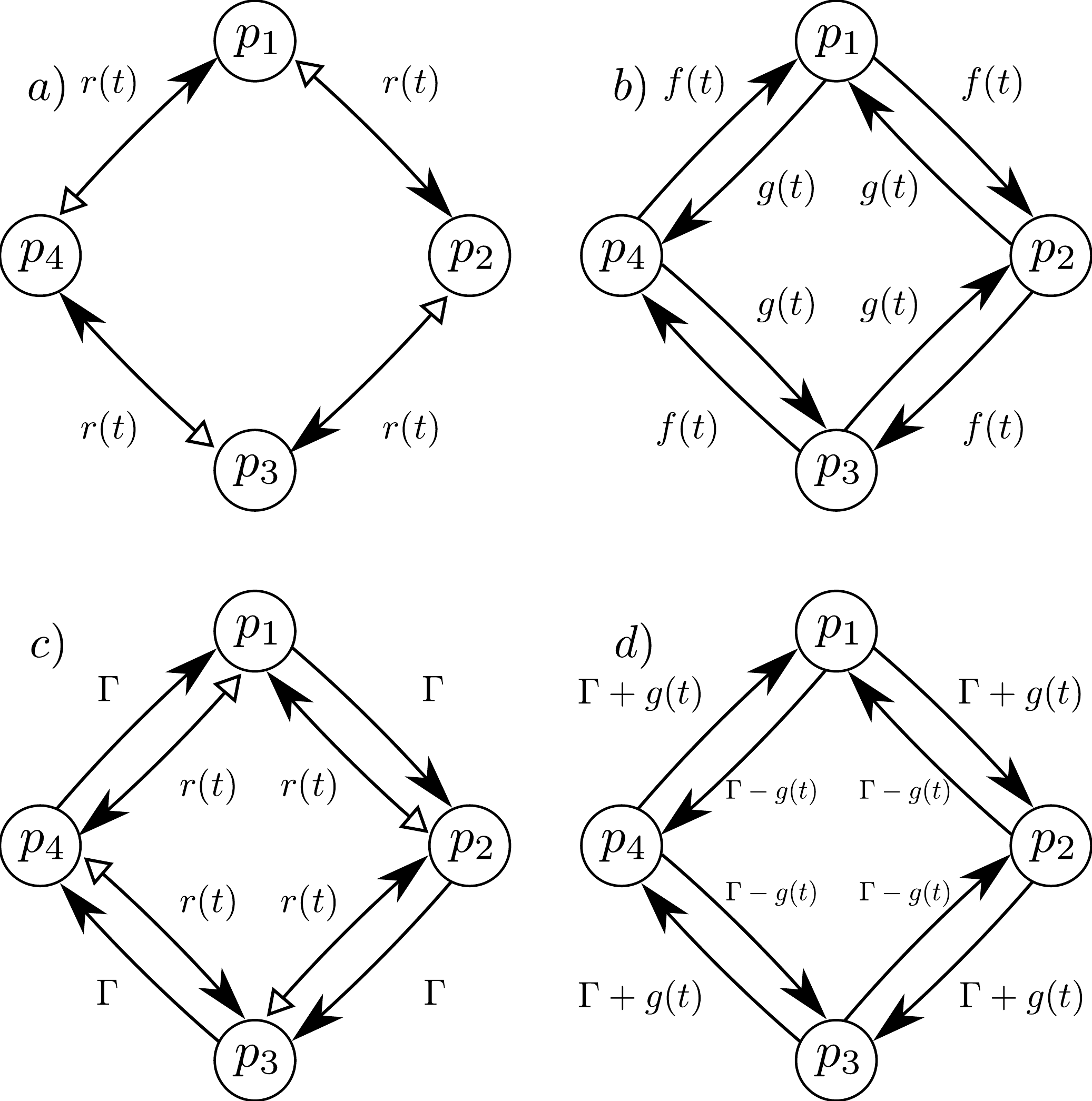}
\caption{\label{fig2} The Markov and non-Markov chains considered in the text. The single headed arrows describe a positive rate and the double headed arrows rates oscillating between positive (black head) and negative (white head) values.}
\end{figure}

A finite continuous time Markov chain is a process ${X(t),t \geq 0}$ with values in ${0,1,2,...,N}$, such that given the present, the past and future are independent, i.e., for all $s,t \geq0$ and all states $i,j,x(u)$ the Markov condition \cite{vanKampen}
\begin{eqnarray}
&&P (X(t + s) = j|X(s) = i, X(u) = x(u), 0 \leq u < s)\nonumber \\
&=& P(X(t + s) = j|X(s) = i)\label{eq14}
\end{eqnarray}
is fullfilled. The Markov property of equation (\ref{eq14}) guarantees, that such a process is described by a rate equation (\ref{eq4}), where all the rates $\gamma_{kl}(t)$ are positive. A straightforward generalization of such a process to the non-Markovian regime can be made by allowing the rates to get temporarily negative values. Naturally, in this case, the condition of equation (\ref{eq14}) is no longer fulfilled. To see some characteristic features of such processes, let us study the ring configurations of figure \ref{fig2}.

We denote the rates for jumps to the clockwise direction as $\lambda_{i,i+1}(t)$. Rates for the jumps to the opposite direction are
denoted by $\mu_{i,i-1}(t)$.
Now, the probabilities $p_i$ are solutions to the following system of ordinary differential equations
\begin{eqnarray*}
  \dot{p}_i(t)=&-p_i(t)\left(\lambda_{i,i+1}(t)+\mu_{i,i-1}(t)\right)\\
  &+p_{i-1}(t)\lambda_{i-1,i}(t)+p_{i+1}(t)\mu_{i+1,i}(t).
\end{eqnarray*}
We study a ring configuration consisting of four states and choose the initial state as $p_1(0)=1$. First we consider the non-Markovian ring configuration presented in figure \ref{fig2}(a).We take the rates to the anti-clockwise direction to be zero and the rates to the clockwise direction to be equal: $\lambda_{i,i+1}(t)=r(t)$ with the rates given by
\begin{equation*}
 r(t)=f(t)-g(t),
\end{equation*}
where
\begin{eqnarray}
f(t)=\frac{1}{2}\Gamma\left(1+
  \mathrm{sgn}[\cos(\Gamma\pi t-\pi/2)]\right), \label{eq15}\\ 
  g(t)=\frac{1}{2}\Gamma\left(1+
  \mathrm{sgn}[\cos(\Gamma\pi t+\pi/2)]\right) \nonumber
\end{eqnarray}
and $\Gamma$ is a positive constant. The rates are periodic and get temporarily negative values and thus the system is a non-Markov chain. 
The dynamics conserves the positivity, which is easy to check, e.g., by exploiting the change of variable
$y_i(t)=\exp\left[ \int_0^t r(s) ds \right] p_i(t)$.  
When $r(t)<0$ the non-Markov chain performs transitions 
in the anti-clockwise direction with the effective rates 
\begin{equation*}
\tilde{\lambda}_{i,i-1}(t)=\frac{p_{i-1}(t)}{p_i(t)}|r(t)|.
\end{equation*}
In the corresponding Markovian case presented in figure \ref{fig2}(b), where the negative rates are replaced by positive rates to the reversed direction, we have $\lambda_{i,i+1}(t)=f(t)$
for the clockwise jumps and $\mu_{i,i-1}(t)=g(t)$ for the anti-clockwise jumps.

\begin{figure}
\includegraphics[width=8cm]{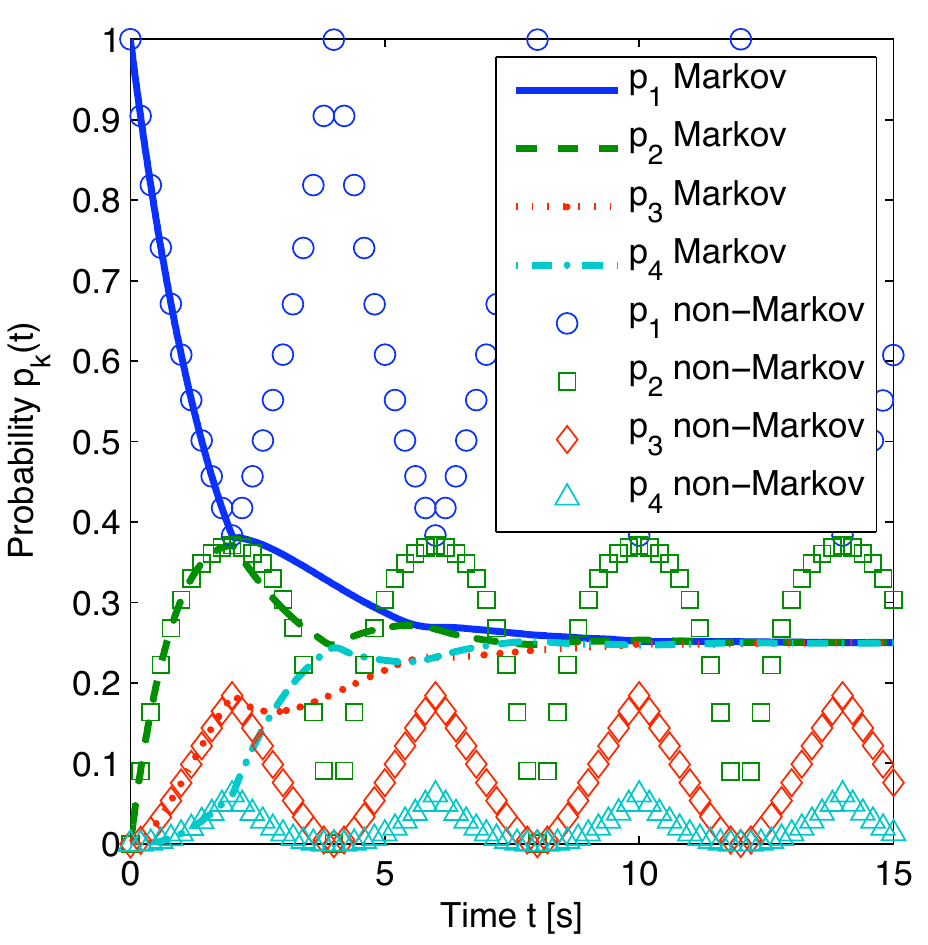}
\caption{\label{fig3} The dynamics of the ring configuration of figures \ref{fig2}(a) (non-Markovian) and \ref{fig2}(b) (Markovian). Initial state for both the Markov and 
   the non-Markov chains is $p_1(0)=1$. $\Gamma=\frac{1}{2}$ 1/s. }
\end{figure}

In figure \ref{fig3} we compare the Markov and the non-Markov chains. We see that the Markov chain has a steady state $p_i=\frac{1}{4}$ which the system reaches independent of the initial state as $t\rightarrow \infty$.The non-Markov chain has a stationary distribution $p_i=\frac{1}{4}$, but since the positive and negative periods have same amplitude and periodicity the non-Markov chain will never reach the stationary state; the periods of negative rate always evolve the system back to the initial state. 

As another example we consider a ring configuration of figure \ref{fig2}(c) with two rates, the one always positive, and the other oscillating between positive and negative values. The initial state 
of the system is $p_1(0)=1$. In the non-Markovian case we have a constant clockwise rate 
$\lambda_{i,i+1}(t)=\Gamma$, and a periodic anti-clockwise rate 
$r(t)$ defined as 
\begin{equation*}
  r(t)=f(t)-g(t),
\end{equation*}
where $f(t)$ and $g(t)$ are defined in equation (\ref{eq15}).
Thus the functional form of the non-Markovian rate is the same as in our previous example but now anti-clockwise. Hence, jumps to anti-clockwise direction take place with rates $\mu_{i,i-1}(t)=\Gamma$ and $\mu_{i,i-1}(t)=0$ periodically while jumps to the clockwise direction occur periodically alternating between  $\Gamma$ and the effective rate $\Gamma+|r(t)|\frac{p_{i+1}(t)}{p_i(t)}=\lambda_{i,i+1}(t)+\tilde{\mu}_{i,i+1}(t)$.

\begin{figure}
\includegraphics[width=8cm]{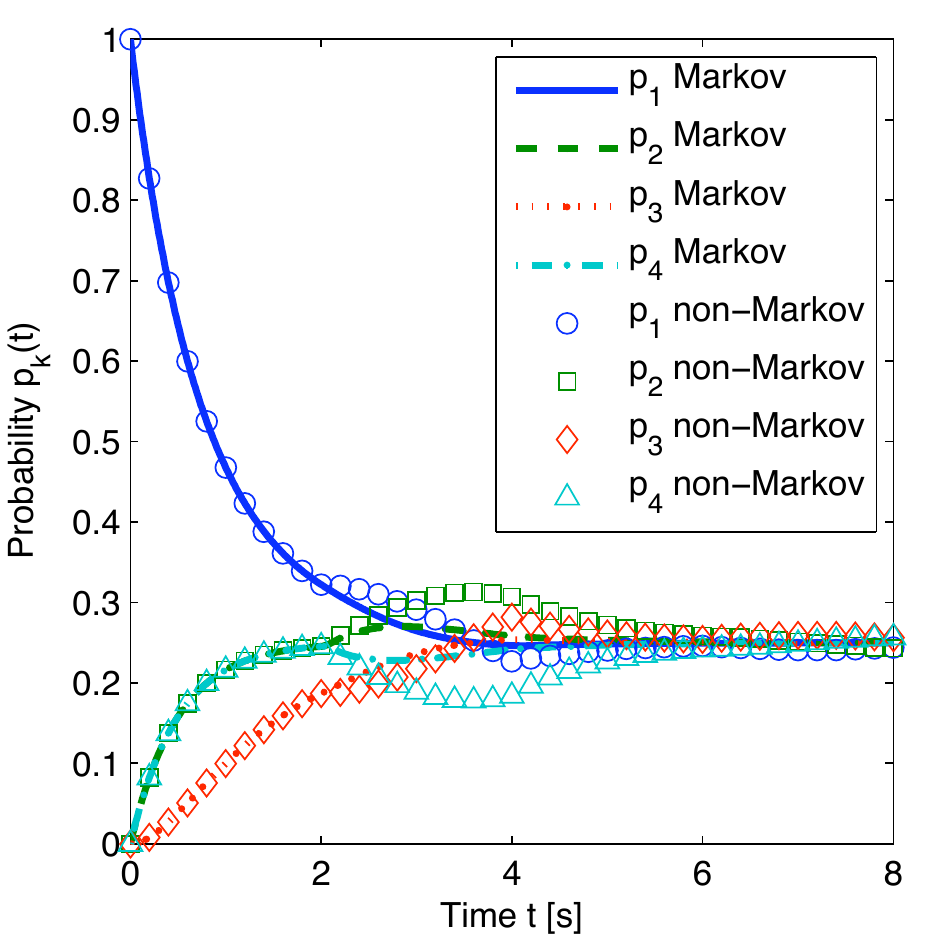}
\caption{\label{fig4} The dynamics of the ring configuration of figures \ref{fig2}(c) (non-Markovian) and \ref{fig2}(d) (Markovian). Initial state for both the Markov and 
   the non-Markov chains is $p_1(0)=1$. $\Gamma=\frac{1}{2}$ 1/s.}
\end{figure}

In the corresponding Markovian case, presented in figure \ref{fig2}(d), we have the rates $\lambda_{i,i+1}(t)=\Gamma+g(t)$ and $\mu_{i,i-1}(t)=\Gamma-g(t)$. The clockwise rate thus takes values $\Gamma$ and $2\Gamma$ and the anti-clockwise rate $\Gamma$ and $0$ periodically. The system has again a stationary state with $p_k(t)=\frac{1}{4}$, which is reached for all initial states as $t\rightarrow \infty$. Also, as we can see in figure \ref{fig4}, the non-Markovian process reaches the stationary state. This is due to two competing processes present in the non-Markovian case: the constantly positive rate drives the system towards the stationary state and the oscillatory rate tries to bring the system towards the initial state. 

\begin{figure}
\includegraphics[width=8cm]{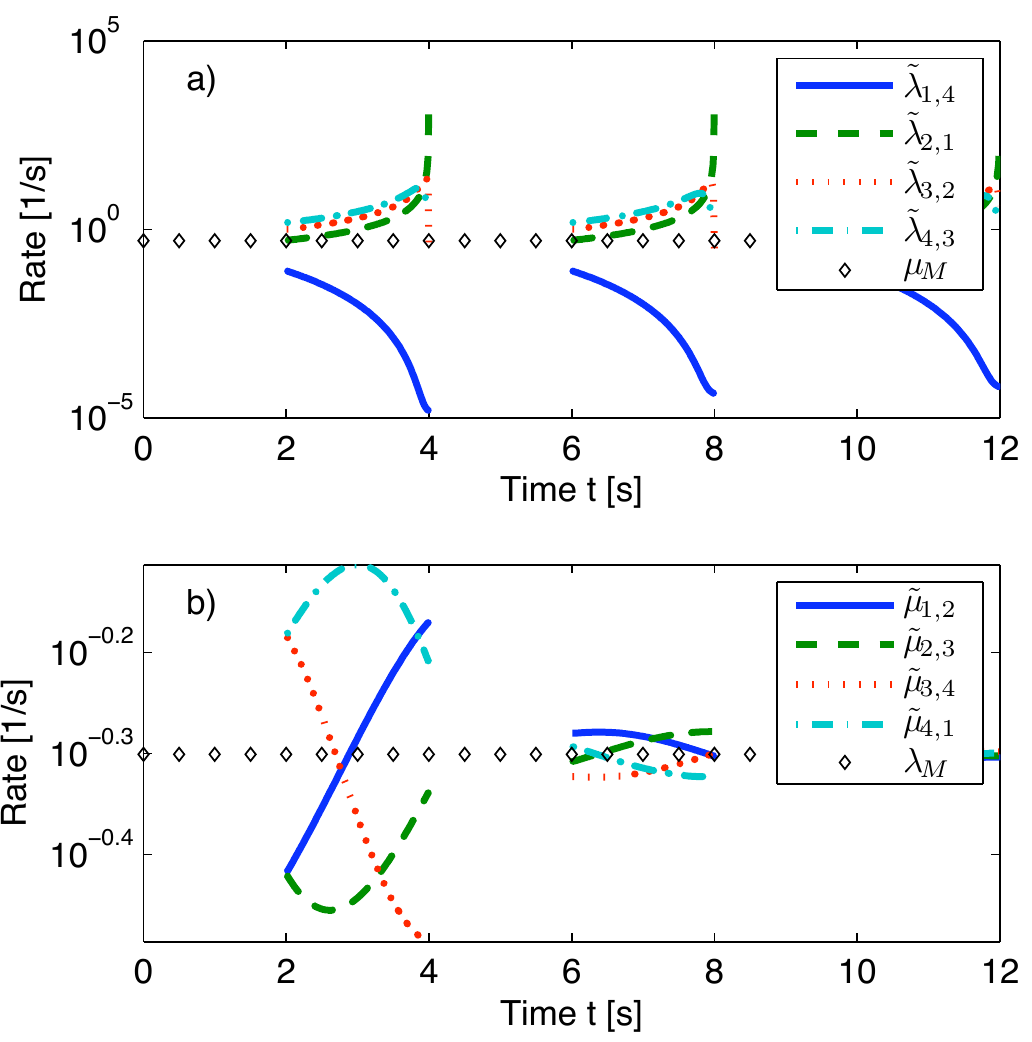}
\caption{\label{fig5} Effective rates of the
non-Markovian ring systems. In panel a) we have the case of 
figure \ref{fig2}(a). In panel b) we have the case of figure  \ref{fig2}(c).}
\end{figure}

The difference in long time behavior of the two non-Markovian processes presented in this section can be seen from the behavior of the effective rates plotted in figure \ref{fig5}. We see that in the first example the effective rates differ greatly from the corresponding Markovian rate independent of how much time has elapsed. In the second example the constantly positive rate drives the effective rates towards the corresponding Markovian value. At later times the effective rates do not differ from the Markovian one. So even though the negative rate is periodic and the magnitude and period do not change in time, the effect of it dies out.

\section{Conclusions} \label{conclusions}
As the need for engineering and controlling open quantum system for quantum information applications has increased in the recent years, efficient methods for treating non-Markovian dynamics are required. The local in time description of non-Markovian dynamics with efficient simulation methods allows a relatively simple treatment for memory effects. Even though the local in time description has been introduced already more than thirty years ago and successfully applied to many non-Markovian problems, nevertheless there has been many misunderstandings related to the applicability of these equations.

In this paper we have discussed these misunderstandings and through examples clarified many aspects related to time local equations. We have also considered classical stochastic processes, which can be derived as a special case of the quantum master equation. We showed that negative decay rates in the master equation do not prevent the complete positivity of the corresponding dynamical map. Further, we demonstrated that the negative decay rates do not only describe a reversal of a process but, also affect the transition probabilities in the process reflected in effective decay rates describing the dynamics.We compared the two cases with positive and negative rates in order to emphasize that local in time equations with negative rates describe memory effects. 

Non-Markovian local in time equations have not yet been extensively used in the context of classical stochastic processes. In this paper we used a local in time description for modeling a non-Markov chain influenced by memory effects. The non-Markovian extension shows many interesting dynamical features absent in the well known Markov chain. We found that the stationary behavior of the process can change drastically, when memory effects are present: instead of reaching a unique stationary state, the system may repeatedly return to its initial state. We expect that application of the local in time description in the classical realm could open new venues of research also in the theory of classical stochastic processes with a variety of applications.

\section*{Acknowledgments}
This work was supported by the Magnus Ehrnrooth Foundation, the Jenny and Antti Wihuri Foundation, the Vilho, Yrj\"o and Kalle V\"ais\"al\"a Foundation, the Graduate School of Modern Optics and Photonics, the Academy of Finland (project 259827), and the COST Action MP1006.

\end{document}